\documentclass[useAMS,usenatbib]{mn2e}
\usepackage{latexsym,graphicx,natbib,amssymb,aas_macros}

\newcommand\msun{\rm\,M_\odot}

\title[gas exp., init. mass-segr. and the st. MF of GCs]
      {The influence of gas expulsion and initial mass-segregation on the stellar mass-function of globular star clusters}
      
\author[Michael Marks, Pavel Kroupa and Holger Baumgardt]
{
  Michael Marks\thanks{e-mail: mmarks@astro.uni-bonn.de (MM);
	holger@astro.uni-bonn.de (HB); pavel@astro.uni-bonn.de (PK)}, 
  Pavel Kroupa and Holger Baumgardt\\
  Argelander Institute for Astronomy, University of Bonn, Auf dem H\"ugel 71, 53121 Bonn,
  Germany\\
}

\begin{document}

\date{Accepted ????. Received ?????; in original form ?????}

\pagerange{\pageref{firstpage}--\pageref{lastpage}} \pubyear{2008}

\maketitle

\label{firstpage}

\begin{abstract}
Recently \citet{dmpp07} studied a sample of twenty globular clusters and found that all clusters with high concentrations have steep stellar mass-functions while clusters with low concentration have comparatively shallow mass-functions. No globular clusters were found with a flat mass-function and high concentration. This seems curious since more concentrated star clusters are believed to be dynamically more evolved and should have lost more low-mass stars via evaporation, which would result in a shallower mass-function in the low-mass part.

We show that this effect can be explained by residual-gas expulsion from initially mass-segregated star clusters, and is enhanced further through unresolved binaries. If gas expulsion is the correct mechanism to produce the observed trend in the $c-\alpha-$plane, then observation of these parameters would allow to constrain cluster starting conditions such as star formation efficiency and the time-scale of gas expulsion.
\end{abstract}


\section{Introduction}
\label{sec:intro}
The stellar initial mass function (IMF) is one of the most fundamental distribution functions of astrophysics. The number of stars that form in one event with stellar masses in the interval $m,m+dm$ is $dN=\xi(m)dm$, where $\xi(m)$ is the IMF. Star-formation theory predicts systematic variations of the IMF with changes of the physical conditions \citep[e.g.][]{ml96,l98,e04,tp05,blz07}. Nevertheless, the IMF has been found to be essentially invariant \citep{man98,maw01,k01,k02}, which is still a challenge for theoreticians.

\begin{figure}
 \centering
 \includegraphics[width=8.3cm]{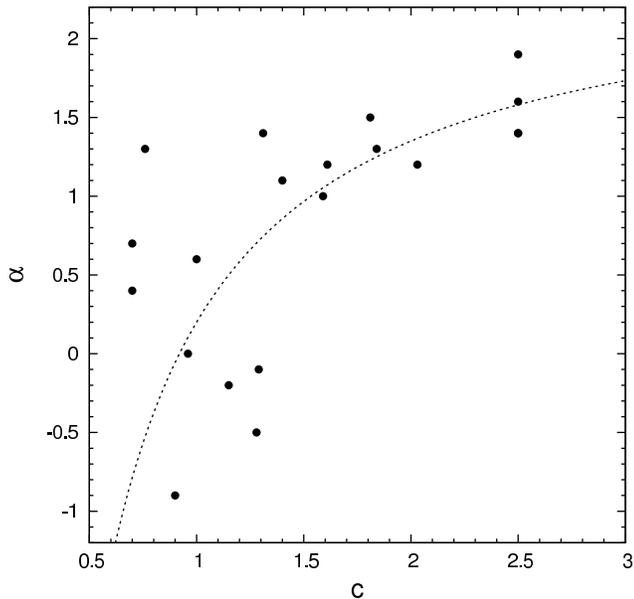}
 \caption{Observed trend between MF index $\alpha$ and the concentration parameter $c$ (taken from fig.1 in DMPP). The dashed line is a fit to the distribution (Section \ref{sec:compobs}). The slope $\alpha$ is measured in the mass-range $0.3-0.8\msun$ and the canonical IMF has $\alpha\approx1.72$ over this mass-range.}
 \label{fig:DMPP}
\end{figure}

\citet*[hereafter DMPP]{dmpp07} reported a surprising relation (see Fig. \ref{fig:DMPP}) between the slope $\alpha$ of the low-mass stellar mass-function $\left(dN/dm=m^{-\alpha}\right)$ of globular clusters (GCs) and their concentration parameter $c=\log\left(r_t/r_c\right)$, i.e. the logarithmic ratio of tidal- and core-radius. All high concentration clusters in their sample had a steep mass-function (MF), while low concentration clusters tended to have a flatter MF (smaller $\alpha$). The canonical IMF has a slope of $\alpha=+1.3$ and $\alpha=+2.3$ in the mass range $0.08\msun-0.5\msun$ and $0.5\msun-150\msun$ \citep*{k01,wk04}, respectively\footnote{\textit{Note:} DMPP defined their MF as $dN/dm=m^{+\alpha}$.}. The mechanism usually believed to be responsible for high cluster concentration is core-collapse driven by two-body relaxation and the same process causes low-mass stars to move to the outer cluster parts where they are removed by the external tidal field. So one would naively expect the relation to be the other way round, and this is indeed confirmed by the direct N-body simulations of initially non-mass-segregated clusters of \citet{bm03}.

Among other approaches to a solution discussed in DMPP, a variation of the IMF, e.g. depending on cluster density,  could possibly solve this problem, if the present day concentration is primordial. This would for the first time be a strong observational evidence for a variable IMF. However, before adopting such an exciting solution, it is important to check for alternative solutions.

Many observations show very young clusters to be (partially) mass-segregated \citep[e.g. recent papers by][]{sabbi07,cdgz07,dib07}. The reaction of initially mass-segregated clusters to gas expulsion and the influence which this has for the stellar MF in a cluster has never been investigated but may be important for understanding the structure and stellar content of present day globular clusters. In this paper we study for the first time the influence of the gas expulsion process on the MFs of initially mass-segregated globular clusters (GCs) with an \textit{in}variable IMF using the recent results by \citet*[see Section \ref{sec:models}]{bk07} for the dynamical response of star clusters to gas expulsion.

The idea behind this is as follows: Star clusters form from massive, dense cores in giant molecular clouds. Stars form in these cores with a given star formation efficiency (SFE) reproducing the stellar IMF. The radiation of the newly born stars will immediately begin to disrupt their surrounding gas. Especially the winds of high-mass (O- and B-type) stars and the first supernovae will lead to ejection and disruption of the entire cloud. The stars have to react to the change in the potential and depending on the gas removal time-scale, the stars will or will not have time to react to this change, so by gas expulsion the final properties of the cluster will be affected. The cluster expands and thereby preferentially stars from the cluster periphery are lost. If mass-segregation is a primordial feature of star clusters, mostly the low-mass stars are lost due to gas expulsion. Thus, the low-mass stellar MF changes strongly, while the high-mass part is left unaffected. Also the concentration is affected by cluster formation parameters such as the SFE and the gas expulsion time. Intuitively we expect the most-disruptive cases of gas expulsion to lead to remnant clusters that have the lowest density and largest depletion of low-mass stars.

The paper is divided in three parts: In Section \ref{sec:models} we explain the models used and the procedure of our analysis. This is followed by a presentation of the results in Section \ref{sec:result} and a discussion \& conclusion in Section \ref{sec:concl}.

\section{The Models}
\label{sec:models}
\citet*[hereafter BK07]{bk07} carried out a large set of $N$-body integrations studying the effect of residual-gas expulsion on the survival rate and final properties of star clusters \citep*[cf.][]{lmd84,kah01}. They placed star clusters on a circular orbit around a spherical galactic potential and varied $(i)$ the star formation efficiency 
\begin{equation}
\epsilon=\frac{M_{\rm ecl}}{M_{\rm ecl}+M_{\rm gas}}\,,
\end{equation}
where $M_{\rm ecl}$ is the stellar mass of the embedded cluster and $M_{\rm gas}$ is the mass of the residual gas within the cluster volume, $(ii)$ the ratio of the gas expulsion time to the crossing time of a cluster, $\tau_M/t_{\rm cross}$, and $(iii)$ the strength of the external tidal field in terms of the initial half-mass radius in units of the tidal-radius $r_h/r_t$. In their models, they assumed that the SFE does not depend on the position inside the cluster, so gas and stars followed the same density distribution initially, which was given by a Plummer model. The influence of the gas on the stars was modelled as a modification to the equation of motion of stars. Gas expulsion was assumed to start at a certain time $t_D$, which was set equal to one $N$-body time unit \citep{hm85}, equivalent to $1/\sqrt{8}$ of a crossing time at the clusters virial (= gravitational) radius \citep{bt87}. After the delay time $t_D$, the gas density was decreased exponentially on a characteristic time-scale $\tau_M$, the gas expulsion time-scale, so the total gas left at later times was given by \citep{kah01}
\begin{equation}
 M_{gas}(t) = M_{gas}(0) \; e^{-(t-t_D)/\tau_M} \; \; .
\end{equation}
All calculations were performed with the collisional $N$-body code NBODY4 \citep{a99} on the GRAPE6 computers \citep{mfkn03} at the Argelander Institute. All clusters contained $20000$ equal-mass stars initially, distributed according to a Plummer sphere. On the one hand, the restriction to equal-mass stars is necessary as otherwise a grid of cluster mass would need to be computed for each of the above parameter combinations. This is presently not feasible. On the other hand, $20000$ stars give good statistics. The integration proceeded for $1000$ initial $N$-body times (equivalent to about $300$ initial crossing times).

The above procedure captures the essence of the physics and their results can be summarised briefly as follows: Both the star formation efficiency and the speed with which the gas is removed have a strong influence on the evolution of star clusters. In the case of instantaneous gas removal ($\tau_M\ll t_{\rm cross}$), clusters have to form with SFEs $\ge 33$ per cent in order to survive gas expulsion \citep{bk03a,bk03b}. This limit is significantly lowered for gas removal on longer time-scales and clusters with SFEs as low as $10$ per cent can survive gas expulsion in the adiabatic limit ($\tau_M\gg t_{\rm cross}$) if the external tidal field is weak. External tidal fields have a significant influence on the cluster evolution only if the ratio of $r_h/r_t$ is larger than about 0.05. Below this value, star clusters behave nearly as if they are isolated.

\subsection{Procedure \& Assumptions}
\label{sec:assum}
For each of the 20000 stars in the models of BK07 we selected a mass from the canonical IMF using a C-routine described in the appendix of \citet*{pk06}. This reproduces the steep MF of the high concentration clusters, which have values around $\alpha\approx1.5$ in the mass range $0.3\msun-0.8\msun$ (hereafter referred to as the low-mass part of the MF) considered by DMPP, if MFs of these clusters still resemble their IMFs\footnote{A power-law fit to the canonical IMF gives $\alpha\approx1.72$ over this mass-range.}. We further omitted the brown dwarf (BD) part of the IMF, i.e. masses below $0.08\msun$, since BDs have a negligible effect on the overall evolution of the cluster and they can't be detected in GCs.

We distributed stellar masses among the 20000 stars according to their initial total energy: The star with the smallest specific energy, i.e. the star that is strongest bound, gets the largest mass and the one with the largest energy is assigned the smallest mass. However, a segregation of 100 per cent as for this mass-assignment is most likeley exaggerated and at least the outer regions of a cluster wouldn't be expected to be totally mass-segregated, so our assumption probes one extreme. We also performed our analysis with initially \textit{not} segregated clusters, where the masses have been assigned randomly to the stars to probe the other extreme. Note again, that the models of BK07 have equal-mass stars, but as the mass in gas in most cases is much larger than the total mass in stars, the dynamical evolution is dominated by the gas expulsion process for the time-span over which the models were computed. Since BK07 included the effect of an external tidal field in the so-called 'near field approximation' \citep*{a85}, the specific energy of a star is given by
\begin{equation}
E_{\rm tot}=\frac{1}{2}\dot{\textbf{\textit{r}}}^2-\Phi\left(\textbf{\textit{r}}\right)-\frac{1}{2}\omega^2\left(3x^2\textbf{\textit{e}}_x+z^2\textbf{\textit{e}}_z\right)\;,
\end{equation}
where the first term is the specific kinetic energy of a star, $\Phi$ is the potential due to the residual-gas and the other stars in the cluster and the last term is a combination of centrifugal and tidal energy. The vector $\textbf{\textit{r}}=\left(x,y,z\right)$ is the position-vector of each individual star measured from the cluster centre, the $\textbf{\textit{e}}_i\;(i=x,\,z)$ are unit vectors and the angular velocity of the cluster around the spherical galactic potential,
\begin{equation}
\omega=\sqrt{\frac{GM_G}{R_G^3}}\;,
\end{equation}
is determined by the mass of the galaxy, $M_G$, and the galactocentric distance, $R_G$.

Recently analytic techniques became available to create mass-segregation in modeled clusters \citep*{gg08,skb08,bkdm08}. \citet{gg08} created a radius dependent 'segregated MF' for the high-mass stars while the low-mass part is radius independent. While our procedure has all massive stars but no low-mass stars in the centre, their formulation of mass-segregation mixes both low and high mass stars in the inner part. Both techniques have no high-mass stars in the cluster outskirts.

In order to determine the final concentration of the modelled clusters, the tidal- and core-radii are needed. In order to determine these, we first calculated the position of the cluster centre (projected centre in the case of the core-radius) using the method by \citet*{ch85}.

The tidal radius of the cluster was then determined iteratively, by first assuming that all stars still in the calculation are bound and calculating the tidal radius,
\begin{equation}
 r_t=\sqrt[3]{\frac{G\,M_{\rm ecl}}{3M_G}}R_G\;,
\label{eq:tidalradius}
\end{equation}
at the end of the the $N$-body runs, when the residual-gas is expelled. In a second step we computed the mass of all stars inside $r_t$ and used it to obtain a new estimate for the tidal radius from equation (\ref{eq:tidalradius}). This was repeated until a stable solution was found.

To identify the core-radius, we placed radial annuli around the projected cluster-centre and searched for the radius at which the projected number-density of stars dropped to half its central value. This method is of course sensitive to the choice of the width of the rings. A too small width leads to a very noisy surface-density profile, too large widths result in inaccurate values for the core-radius. Also the lower the number of stars at the end of the computations the noisier the surface-density-profile gets. So we decided to increase the ring-width for clusters with less than 5000 (25 per cent) and 2000 (10 per cent) stars at the end of the $N$-body integrations, respectively, in order to compensate for that effect. This hardly changes the measured core-radii for smooth density-profiles, but improves them significantly for those with a not so stable profile. The central-density was determined by averaging over the surface-densities in the first two to five annuli (depending on the ring-width) to reduce the influence of outliers. We took the mean value of the measured core-radii from three different projections as the best approximation of the true core-radius in the models. Their standard-deviation was taken to be the error of $r_c$. Models which survived gas expulsion but kept less than five per cent of their stars have been left out of consideration because radii-determinations were too difficult.

With the tidal- and core-radius, the concentration and its error follow from
\begin{equation}
c=\log_{10}\left(\frac{r_t}{r_c}\right)
\end{equation}
and error propagation.

\begin{figure*}
 \begin{center}
  \includegraphics[width=17cm]{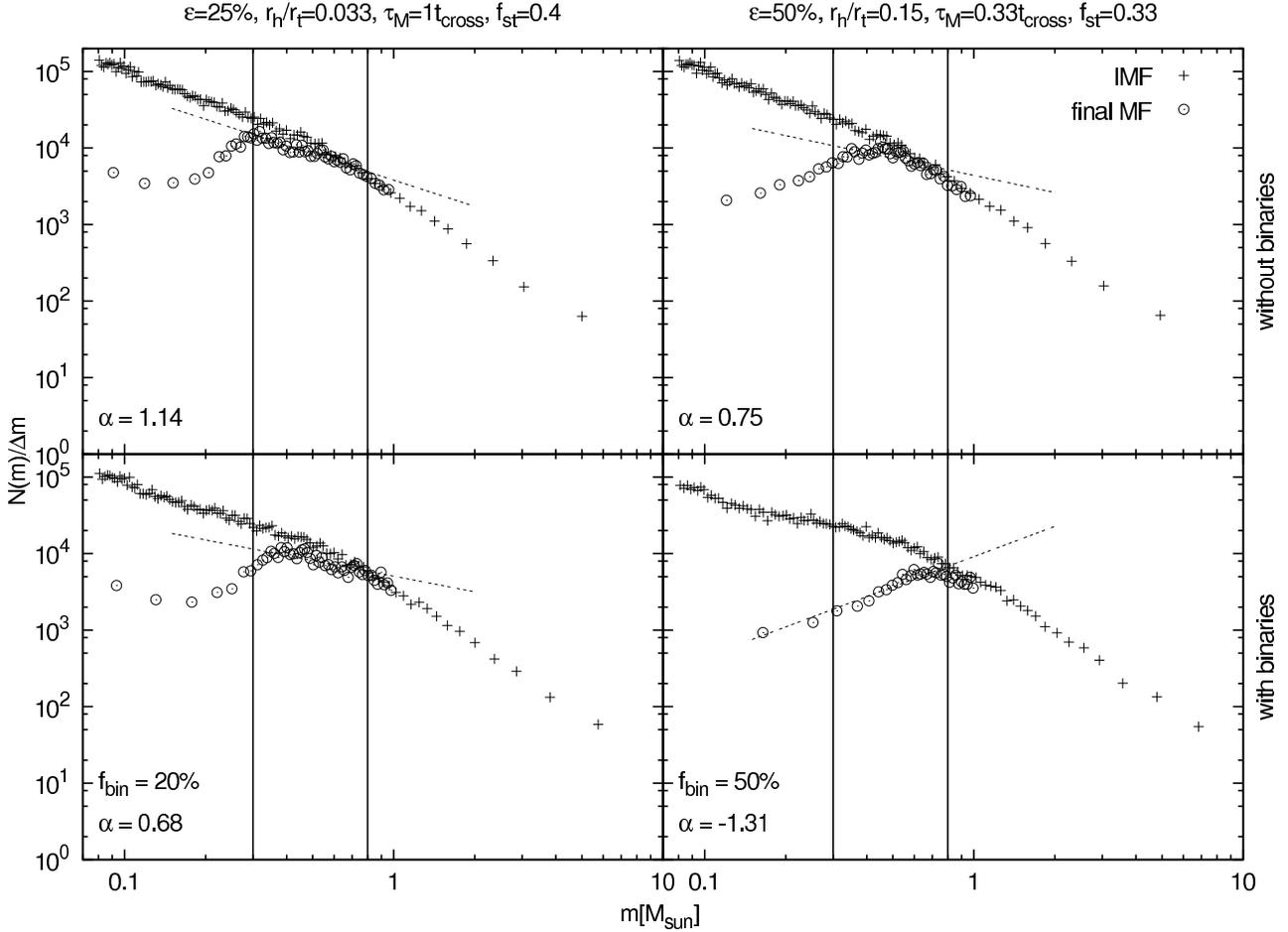}
 \end{center}
 \caption{Example mass-spectra for our analysis. The plots show the number of stars, $N(m)$, per mass-interval normalised to the bin-width, $\Delta m$, per bin versus the mass $m$, both on logarithmic scales. The left and right panels, respectively, show the same models (see title), one without (upper) and one with unresolved binaries (lower). The fraction of stars remaining in the cluster is $f_{\rm st}$. Crosses indicate the IMF, open circles the resulting MF. The solid vertical lines limit the mass range $0.3-0.8\msun$. The dashed line is the fit to the final MF in this region. In mass-segregated clusters only the low mass-part of their MF is affected: The slope of the MF shifts to lower values for $\alpha$ if binaries are included in our analysis. Also note that the MFs with unresolved binaries are flatter for larger binary-fractions \citep*[cf.][]{kgt91}.}
 \label{fig:massspectra}
\end{figure*}
Furthermore, at the end of each integration, we plotted the MF of the stars that lie within the final tidal radius following the binning-method described by \citet*[their experiment 3]{ma05} assuming a power-law behaviour in the interesting mass-range: We chose to distribute the masses in $2\times N^{2/5}$ bins \citep*[second recommendation of][]{ds86} with an equal number of stars in each bin\footnote{Binning was performed over the whole mass-range $0.08\ldots150\msun$.}. Afterwards we determined the slope $\alpha$ of the resulting MF in the mass range $0.3\msun-0.8\msun$ by applying a least-squares fit. Although the slope may change over this range (see upper right panel of Fig. \ref{fig:massspectra}), we assigned just one $\alpha$-value for each model. This may not be completely representative for the actual mass-distribution for some models, but the uncertainties of the best-fitting line give an idea of how strong the data deviates from an ideal power law. The errors (indicated in Fig. \ref{fig:comparison}) can be seen as the uncertainty in the determination of the slope.

\section{Results}
\label{sec:result}
\subsection{The effect of gas expulsion on the stellar MF}
\label{sec:MFgas}
\begin{figure*}
 \begin{center}
  \includegraphics[width=17cm]{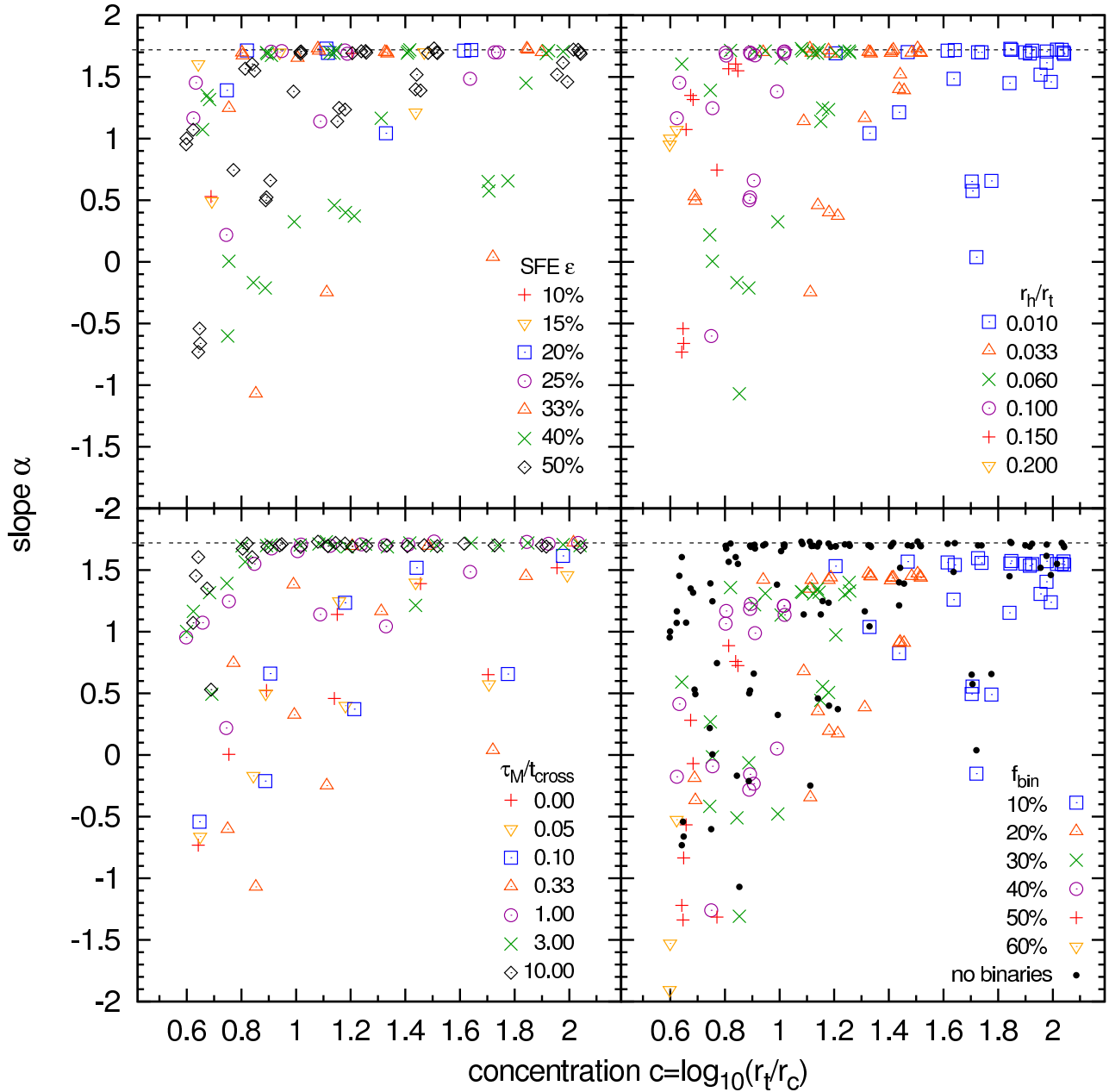}
 \end{center}
 \caption{The slope $\alpha$ of the MF at the end of the $N$-body integrations in the mass-range $0.3\msun-0.8\msun$ as a function of the cluster concentration $c$. The upper panels and the lower left show the same data, but coded for a different parameter used in the set of models of BK07: \textit{top left:} SFE $\epsilon$; \textit{top right:} strength of the tidal field measured by the ratio of $r_h/r_t$; \textit{bottom left:} for the characteristic gas expulsion-time-scale, $\tau_M$, measured in crossing-times, $t_{\rm cross}$, of the initial cluster. The horizontal dashed line gives the slope of the canonical IMF in the mass-range $0.3\msun-0.8\msun$; all initial models lie on this line and have $1\lesssim c_{\rm init}\lesssim 2.5$. The lower right plot compares the position of the points in the c-alpha-plane with (open symbols and crosses) and without (filled dots) binaries in dependence of the chosen binary-fraction. For the sake of clarity, error bars are omitted in these plots. Uncertainties are indicated in Fig. \ref{fig:comparison}}
 \label{fig:calpha}
\end{figure*}
Fig. \ref{fig:calpha} depicts the effect of gas expulsion on the slope of the MF in the low-mass part in dependence of the cluster final concentration. As can be seen, the MFs of most high-concentration clusters still resemble their IMFs and don't show a strong change of the slope in the low-mass part, while for clusters with low-$c$ several models do show a strong change in the slope. The exact position of a point in the diagram depends on the set of parameters for the corresponding model.

In the upper part of the diagrams (above $\alpha\approx1$) one finds the models which kept most of their stars. These are preferentially models with a gas expulsion time-scale $\gtrsim t_{\rm cross}$. Except for the lowest concentration clusters, most of them still lie on their IMF slope. Those models with a smaller $\tau_M$ therefore have an extremely high SFE (above 40 per cent) and experience a weak or intermediate tidal field only. Just three models with a gas expulsion time-scale larger or equal to the cluster's initial crossing time are found to be shallower than $\alpha\approx-1$ which is mainly due to their low SFE of $\lesssim15$ per cent. Below a concentration of $c\sim0.8$ the strong tidal fields in these models lift even the large-$\tau_M$ clusters with a large SFE away from the IMF slope.

In the left part of the diagrams ($c\lesssim1.4$) several models show a strongly flattened MF (below $\alpha\approx1$), which are predominantly the models with fast gas removal ($\tau_M\ll t_{\rm cross}$). These clusters experience intermediate or strong tidal-fields and have large SFEs ($\gtrsim33$ per cent), which is responsible for cluster survival.

The upper right panel reveals the dependency of the final concentration on the the initial tidal field strength. Starting almost exclusively with weak tidal fields at large concentrations, stronger tidal fields become successively more common as the concentration decreases. This is easy to understand: Due to general cluster expansion the tidal boundary of the cluster shrinks and the core radius grows, the final concentration can be expected to be lower for models in which the two radii are lying closer together initially, i.e. for the clusters in a stronger tidal field. Of course the final concentration is also affected by the gas expulsion time and the SFE, since they determine how large cluster expansion exactly is. That is, the models with different tidal field strength overlap slightly, but the overall $c$-value is established by the tidal field. Strong tidal fields at large galactocentric distances may have occurred for those GCs forming in the immediate vicinity of pre-Milky Way gaseous building blocks.

\begin{figure*}
 \begin{center}
  \includegraphics[width=17cm]{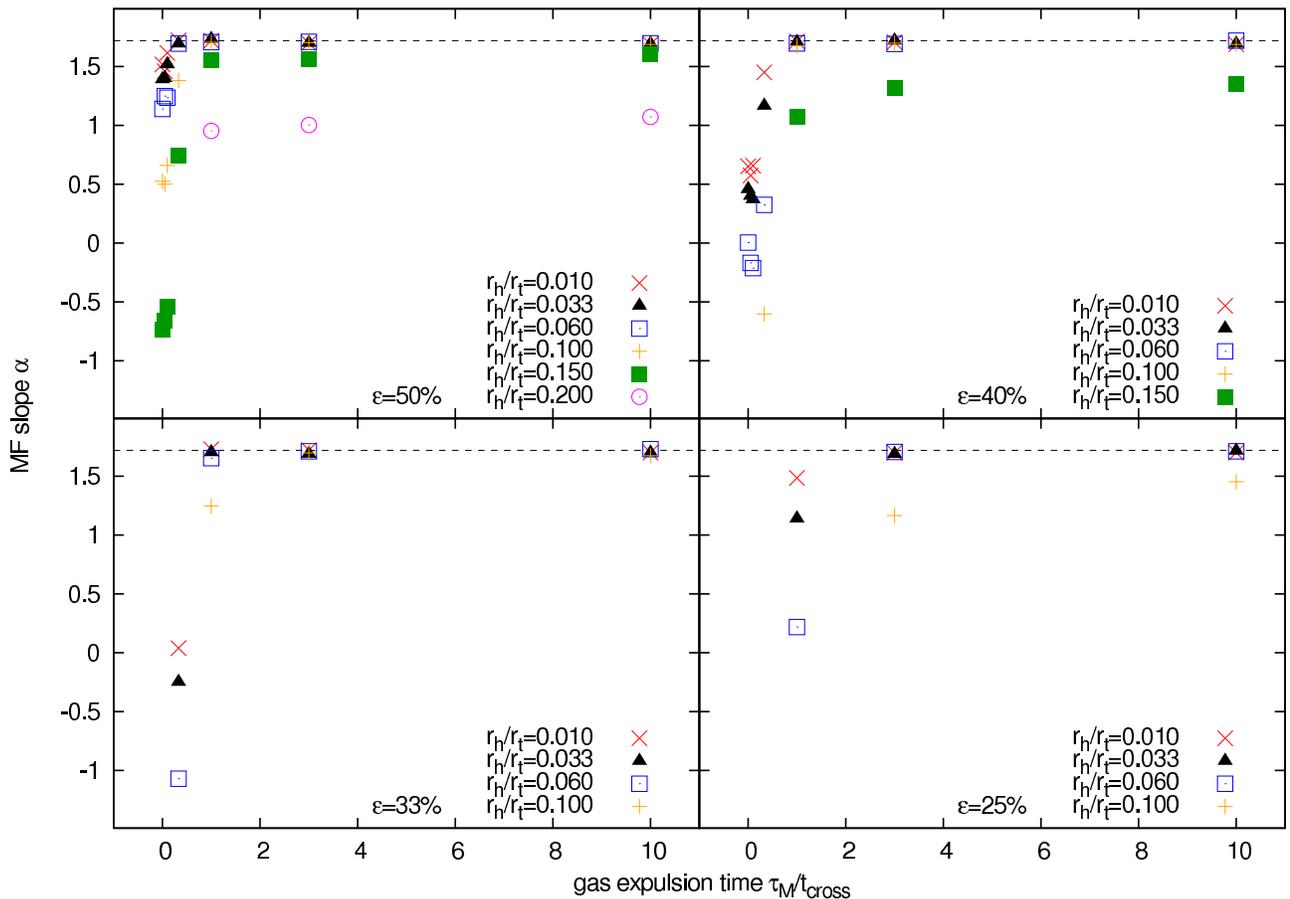}
 \end{center}
 \caption{Dependence of the slope $\alpha$ of the low-mass stellar MF (as in Fig. \ref{fig:calpha}) on the gas expulsion time $\tau_M$ for a fixed value of the SFE $\epsilon$ (as indicated in the plots). The horizontal dashed line shows the IMF slope.}
 \label{fig:alphatgas}
\end{figure*}
The lower left panel in Fig. \ref{fig:calpha} suggests a systematic behaviour of the MF slope with the concentration for a given gas expulsion time-scale. This can explicitly be seen for $\tau_M\geq t_{\rm cross}$. The panels of Fig. \ref{fig:alphatgas} emphasize the strong dependence on this parameter for fixed SFEs. As the gas expulsion time decreases the MF becomes shallower. Furthermore, a stronger tidal-field leads to enhanced mass-loss and thus to a lower MF-slope. So the models with the strongest tidal-fields are the ones with the lowest value of $\alpha$ for a fixed gas expulsion time. The lower the SFE gets, the stronger is the effect of the gas expulsion time. For a high SFE, very strong deviations from the IMF slope start for the lowest gas expulsion times and the strongest tidal-fields first, migrating to larger $\tau_M$ and lower $r_h/r_t$ as the SFE decreases.

The influence of the exact set of parameters is so strong, that for gas expulsion times $\lesssim0.33\,t_{\rm cross}$ we even observe four strongly affected MFs in the high-concentration regime.

\subsection{Unresolved binaries}
\label{sec:unrbin}
An error occurring in any observational determination of MF slopes is due to unresolved binaries. \citet*{kgt91} and \citet{k01} showed that this leads to an additional apparent flattening of the MF. We tried to account for this fact in our analysis as follows: We randomly picked stars from the stellar mass grid and randomly chose a secondary mass for them from the canonical IMF (random pairing). A 'centre-of-mass' star is then assigned the sum of the two selected masses. These 'binaries' are still treated as single stars, i.e. they are 'unresolved', but they now have the combined mass of their components. So the companions will stay together throughout the simulation and stay bound to or leave the cluster as a stellar system, i.e. possible hardening effects or dissolution of binaries is not directly being considered. But it is included indirectly by assuming a variable binary-fraction in dependence of $c$.

The binary-fraction of a cluster is defined as
\begin{equation}
f_{\rm bin}=\frac{\rm number\;of\;binary\;stars}{\rm number\;of\;single\;stars+binary\;stars}\;.
\end{equation}
In order to account for a change of the binary-fraction of a cluster due to dynamical evolution, we vary $f_{\rm bin}$ for different models. In denser clusters close encounters between stars are more common than in less dense models \citep*{k95}. So dissolution of binary- and multiple-systems is more likeley in denser environments. Assuming that star clusters form with similar binary-fractions and that concentration is a measure of cluster density\footnote{Depending on the actual value of $r_t$ and $r_c$: in principle a cluster with larger $c$ can be less dense than a cluster with lower $c$.}, clusters with a high concentration after gas expulsion are expected to retain a smaller number of binaries than low-concentration clusters after sufficient time for dynamical evolution. As has been shown in Section \ref{sec:MFgas}, the concentration at the end of our $N$-body integrations approximately follows the strength of the tidal field. So we chose to distribute binary-fractions of $f_{\rm bin}=10\%,\;20\%,\;$\ldots$60\%$ according to models with tidal-field strength $r_h/r_t=0.01,\;0.033,\;\ldots0.2$, respectively. Thus, clusters experiencing a weaker tidal field are given a smaller binary-fraction and vice-versa.
\begin{figure*}
 \begin{center}
  \includegraphics[width=17cm]{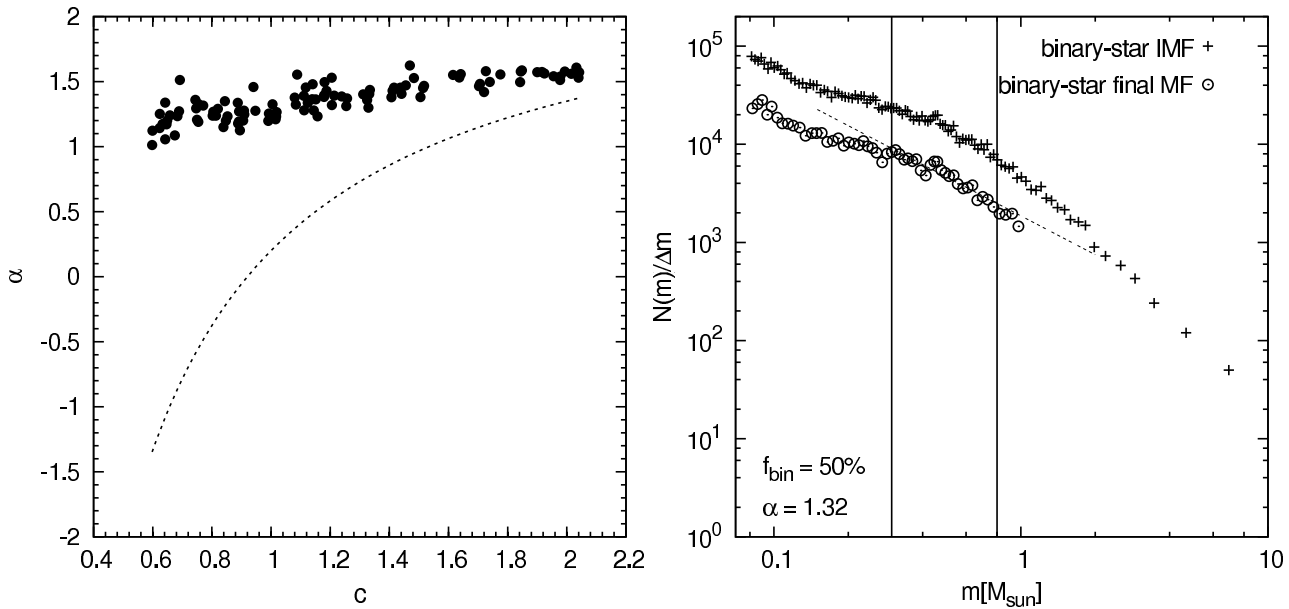}
 \end{center}
 \caption{\textit{Left panel:} The $c-\alpha$ plane for initially non-mass-segregated clusters with unresolved binaries. The dashed line is the relation proposed by DMPP (see Section \ref{sec:compobs}). Just a slight decrease of the slope can be seen due to enhanced binary-fractions in low-concentration clusters. \textit{Right panel:} resulting mass-spectrum (as in Fig. \ref{fig:massspectra}) for no mass-segregation assuming unresolved binaries for the strongly affected cluster in the mass-segregated case with parameters $\epsilon=50\%,\;\tau_M=0.33t_{\rm cross},\;r_h=0.15r_t$ (compare with lower right panel of Fig. \ref{fig:massspectra}). The shape of the MF stays more or less the same, i.e. all stars are equally affected by gas expulsion.}
 \label{fig:masssegr}
\end{figure*}

A choice of $f_{\rm bin}$ independent from the concentration is not suitable to enhance the trend, because the strength of the shift depends on the binary-fraction. In principle one could explain the relation with unresolved binaries alone, if the true MFs of all clusters still resemble their IMF. But since GCs are old objects and should therefore be at least partially dynamically evolved, this is not expected for all of them (Section \ref{sec:intro}). For our modelling, gas expulsion is in any case needed to initiate the trend and to span the whole range in concentration.

After selecting the binaries and assigning masses we distributed these stellar masses to the stars as described in Section \ref{sec:assum} and re-ran our analysis.

As expected, unresolved binaries cause an additional shift of the MF slope. The effect can be seen in the mass-spectra of Fig. \ref{fig:massspectra} (comparison of upper and lower panels) and its influence on the $\alpha-c$-plane is shown in the lower right panel of Fig. \ref{fig:calpha}. The slope of the MF changes according to the chosen binary-fraction towards lower values for $\alpha$ when compared with the mass-spectra or the location of the points in the $\alpha-c$-plots, that consider gas expulsion only. The effect becomes stronger the larger the binary-fraction $f_{\rm bin}$ is. By our assumption, the strongest change in the slope of the MF is seen in the least concentrated final models, because these have been assigned the largest binary-fractions. All models with binaries are flatter than their IMF. The binary MF slopes differ from the non-binary MF slopes by $\Delta\alpha=\left|\alpha_{\rm bin}-\alpha_{\rm no-bin}\right|=0.005$ (one of the $f_{\rm bin}=10$ per cent models) to $\Delta\alpha=2.8$ ($f_{\rm bin}=60$ per cent) with an average change of $\Delta\alpha=0.5$.

\subsection{Not-mass-segregated models}
\label{sec:masssegr}
Fig. \ref{fig:masssegr} (left panel) shows that primordial non-mass-segregated clusters do not show a strong trend of the MF slope with cluster concentration. Just a slight decrease of $\alpha$ is seen as $c$ becomes smaller, which is due to the larger binary-fractions assigned for low $c$-values.

The reason for this behaviour can nicely be seen in the mass-spectrum shown in the same figure (right panel). The diagram displays the same model parameters as Fig. \ref{fig:massspectra} (lower right panel), but for the case of no mass-segregation. Here, the MF is just shifted along the y-axis while the slope hardly changes. Hence, stars of all stellar masses are equally removed from a non-mass-segregated cluster during expansion following gas expulsion leaving $\alpha$ unchanged.

\subsection{Comparison with observations}
\label{sec:compobs}
\begin{figure*}
 \begin{center}
  \includegraphics[width=17cm]{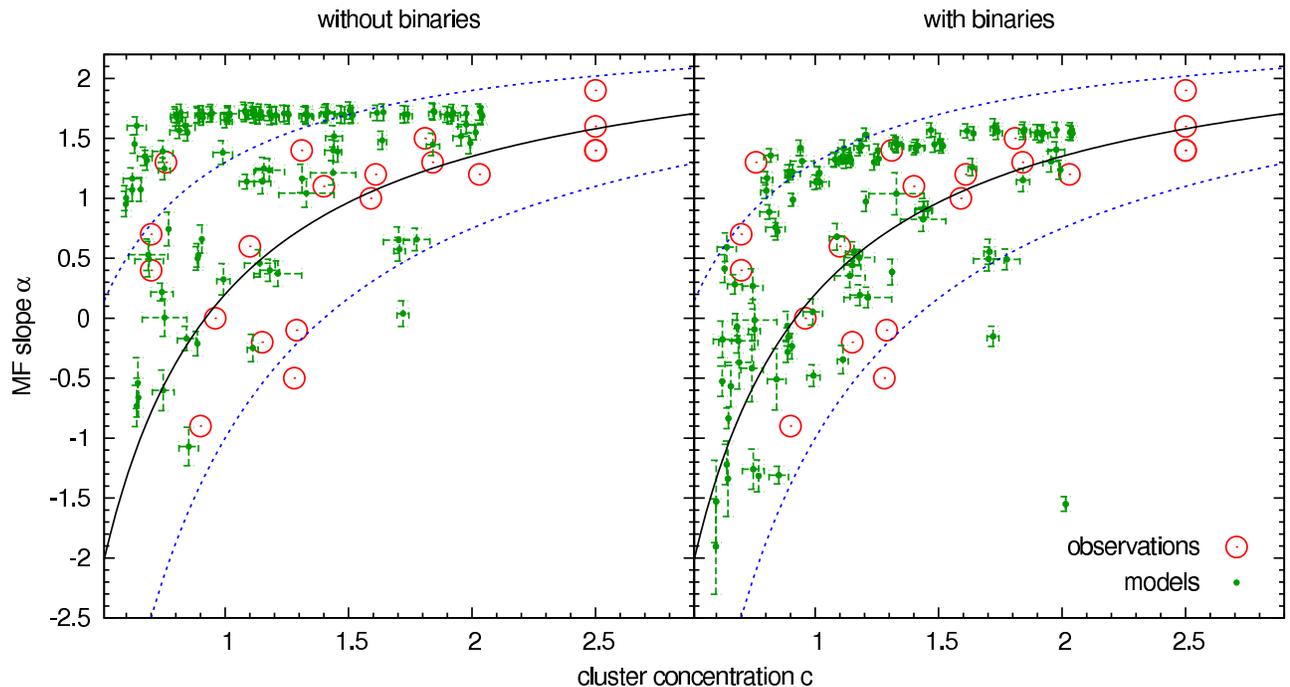}
 \end{center}
 \caption{Comparison of our results with the observational data by DMPP. The left panel shows the result considering gas expulsion only, the right one includes primordial binaries. The large open circles are the observed clusters, the small filled dots (with error bars) are from the $N$-body integrations. The solid line is an eye-ball fit from DMPP (eq. (\ref{eq:eyeball})); we added the dashed lines as limits to their observations (see text). The agreement is better for the models including the effect of unresolved binaries.}
 \label{fig:comparison}
\end{figure*}
The panels of Fig. \ref{fig:comparison} compare the influence of gas expulsion on the MFs of mass-segregated clusters with and without unresolved binaries, with the DMPP data. Fitting by eye, DMPP proposed a relation of the form (solid line)
\begin{equation}
\alpha\left(c\right)=-\frac{2.3}{c}+2.5\,.
\label{eq:eyeball}
\end{equation}
In this Figure we added upper and lower limits (dashed lines) to their observational data
\begin{equation}
\alpha_u\left(c\right)=-\frac{1.2}{c}+2.5\;,\\
\alpha_l\left(c\right)=-\frac{3.5}{c}+2.5\,.
\end{equation}
Both, the models with and without unresolved binaries, agree with the general trend that less concentrated clusters show a shallower MF, but the agreement is much better for the models including unresolved binaries. Even high-$c$ clusters with a flattened MF are within or near the observational limits. The not mass-segregated models, on the other hand, aren't able to reproduce the observed trend (Fig. \ref{fig:masssegr}). Unfortunately we do not reach values around $c\sim2.5$ with our models, but we wouldn't expect significant deviations from the IMF there. Nevertheless, our data points suggest a slightly steeper run than that proposed by DMPP, which can be understood, because DMPP had reliable observational data for only 20 GCs and the relation was an eye-ball fit. In addition, the N-body models of BK07 did not include the effect of dynamical evolution of the clusters beyond the 300 initial crossing times (Section \ref{sec:models}), which will also change the MF slope because of low-mass star evaporation \citep*{bm03}. Dynamical evolution, however, also affects the concentration and the core-collapse process may produce concentrations as large as in the observed clusters with $c\sim2.5$. The difference between the models and observed clusters may also be explained by the exact number of unresolved binaries. How clusters with different initial stellar content and concentration evolve exactly after gas expulsion is still an open issue.

The low-concentration model clusters in the upper left corner of Fig. \ref{fig:comparison} with only one observed cluster, both with and without binaries, are expected to shift to lower slopes during dynamical evolution. Such low-$c$ clusters can indeed survive for a Hubble time and wander downwards in the $c-\alpha$-plane to slightly larger concentrations and join the strongly affected clusters before they dissolve, if these clusters start mass-segregated \citep*[in prep.]{bkdm08}. The low-$c$ clusters with a short gas expulsion time-scale, which expanded a lot and lost many low-mass stars, may be left in a collision-less state such that dynamical evolution doesn't affect the properties of these clusters much.

Although we find several models (including binaries) with MF slopes $-1\gtrsim\alpha\gtrsim-2$ in the lowest concentration regime, which follow the DMPP relation, they do not observe any clusters there. This may be understood because most of the computed clusters in this domain experience the strongest tidal fields and these models are unlikely to survive for a Hubble-time as they expand by an additional $\sim30$ per cent due to stellar evolution and will probably dissolve rather quickly \citep*{fh95,bm03}.

The four model clusters at $c\sim1.7$, which are more strongly depleted in low-mass stars than the other models in that part of the diagram are also further away from the observational data, although three of them are still within the observational limits set by us. This may be a selection effect, because we find just four out of 31 clusters above $c=1.4$ with a strong effect on the MF slope and DMPP observed just six clusters in total in that regime.

Note that we measured global MFs (GMF) from our integrations while in observations only local MF measurements are available, which are subsequently transformed to GMFs by assuming that the MF at the half-mass radius approaches closely the GMF \citep*{dmpp00,bm03}. We checked this in our N-body runs by looking at the local MF (LMF) of the stars near the half-mass radius, i.e. between the 40 and 60 per cent Lagrange radii, and found that we can reproduce the trend as well. The mean difference between LMFs and GMFs in our analysis is $\Delta\alpha=\left|\alpha_{\rm GMF}-\alpha_{\rm LMF}\right|\approx0.25$. This difference might be removed by dynamical evolution.

\section{Summary \& Conclusions}
\label{sec:concl}
We have for the first time studied the effect of gas expulsion with and without unresolved binaries on the stellar MF of initially mass-segregated clusters with an invariable (canonical) IMF.

Due to gas expulsion the cluster expands and stars in the cluster halo are more easily lost than stars near the centre. In a mass-segregated cluster this leads to the loss of the less massive stars causing a change of the low-mass MF. Our computations showed a flattening of the MFs in the low-mass part for less concentrated final clusters, but not for strongly concentrated clusters. This can explain the \citet*{dmpp07} data, while initially non-mass-segregated models don't produce the observed trend. Depending on the parameters of the models, the final concentration and the slope of the MF are determined. Primordial binaries induce an additional apparent shift in the MF slopes (compare panels in Fig. \ref{fig:comparison}), which lead to an even better agreement with the DMPP data. If the difference between theory and observation is due to unresolved binaries, adjusting the exact value of $f_{\rm bin}$ of the model clusters to best reproduce the observational data could lead to predictions of binary-fractions for the observed clusters.

The models can approximately be divided into two different parts: The models with a high concentration of which most still have their IMF, and clusters with a low concentration, where a few systems lost a large portion of their (preferentially) low-mass stars and thus show a significant deviation from their IMF slope.

Our models constitute a possible solution to the DMPP problem without invoking a variable IMF. This contribution emphasizes the importance of choosing physically realistic initial models for understanding cluster evolution: The ability of our models to reproduce the observed $c-\alpha$ trend strongly suggests that initial mass-segregation and gas expulsion play a critical role in star-cluster evolution and that we can't exclude the possibility of a universal IMF for all globular clusters.

However, the MF slope and concentration in our calculations are measured shortly after the residual-gas has been expelled from the cluster once the surviving clusters have re-virialized. Later dynamical evolution of initially mass-segregated clusters will additionally affect these two parameters. This topic is investigated in \citet*[in prep.]{bkdm08}.

If gas expulsion is the determining factor for the observed trend between $c$ and $\alpha$ then, by direct measurement of these parameters, observers may be able to constrain starting conditions of the clusters such as the SFE and the gas expulsion time-scale from Fig. \ref{fig:calpha}, unless the $c$-$\alpha$-plane is degenerate. From our results it is not clear if the slope of the MF is uniquely determined by just one set of parameters for a given concentration. A different combination of the three parameters may yield a similar MF slope (but also the concentration is affected by them). This is especially true for clusters with less affected MFs, i.e. large $\tau_M$ and SFE, but also in other parts of the diagram some data points accumulate. In the case of ambiguous results, at least another parameter would be needed to lift this degeneracy.

\bibliographystyle{aa}
\bibliography{biblio}

\begin{thebibliography}{36}
\expandafter\ifx\csname natexlab\endcsname\relax\def\natexlab#1{#1}\fi

\bibitem[{{Aarseth}(1985)}]{a85}
{Aarseth}, S.~J. 1985, in Multiple time scales, p. 377 - 418, 377--418

\bibitem[{{Aarseth}(1999)}]{a99}
{Aarseth}, S.~J. 1999, \pasp, 111, 1333

\bibitem[{{Baumgardt} {et~al.}(2008){Baumgardt}, {de Marchi}, \&
  {Kroupa}}]{bkdm08}
{Baumgardt}, H., {de Marchi}, G., \& {Kroupa}, P. 2008, \apj, in preparation

\bibitem[{{Baumgardt} \& {Kroupa}(2007)}]{bk07}
{Baumgardt}, H. \& {Kroupa}, P. 2007, \mnras, 380, 1589

\bibitem[{{Baumgardt} \& {Makino}(2003)}]{bm03}
{Baumgardt}, H. \& {Makino}, J. 2003, \mnras, 340, 227

\bibitem[{{Binney} \& {Tremaine}(1987)}]{bt87}
{Binney}, J. \& {Tremaine}, S. 1987, {Galactic dynamics} (Princeton, NJ,
  Princeton University Press, 1987, 747 p.)

\bibitem[{{Boily} \& {Kroupa}(2003{\natexlab{a}})}]{bk03a}
{Boily}, C.~M. \& {Kroupa}, P. 2003{\natexlab{a}}, \mnras, 338, 665

\bibitem[{{Boily} \& {Kroupa}(2003{\natexlab{b}})}]{bk03b}
{Boily}, C.~M. \& {Kroupa}, P. 2003{\natexlab{b}}, \mnras, 338, 673

\bibitem[{{Bonnell} {et~al.}(2007){Bonnell}, {Larson}, \& {Zinnecker}}]{blz07}
{Bonnell}, I.~A., {Larson}, R.~B., \& {Zinnecker}, H. 2007, in Protostars and
  Planets V, ed. B.~{Reipurth}, D.~{Jewitt}, \& K.~{Keil}, 149--164

\bibitem[{{Casertano} \& {Hut}(1985)}]{ch85}
{Casertano}, S. \& {Hut}, P. 1985, \apj, 298, 80

\bibitem[{{Chen} {et~al.}(2007){Chen}, {de Grijs}, \& {Zhao}}]{cdgz07}
{Chen}, L., {de Grijs}, R., \& {Zhao}, J.~L. 2007, \aj, 134, 1368

\bibitem[{{D'Agostino} \& {Stephens}(1986)}]{ds86}
{D'Agostino}, R.~B. \& {Stephens}, M.~A. 1986, {Goodness-of-fit techniques}
  (Statistics: Textbooks and Monographs, New York: Dekker, 1986, edited by
  D'Agostino, Ralph B.; Stephens, Michael A.)

\bibitem[{{De Marchi} {et~al.}(2000){De Marchi}, {Paresce}, \&
  {Pulone}}]{dmpp00}
{De Marchi}, G., {Paresce}, F., \& {Pulone}, L. 2000, \apj, 530, 342

\bibitem[{{De Marchi} {et~al.}(2007){De Marchi}, {Paresce}, \&
  {Pulone}}]{dmpp07}
{De Marchi}, G., {Paresce}, F., \& {Pulone}, L. 2007, \apjl, 656, L65

\bibitem[{{Dib} {et~al.}(2007){Dib}, {Shadmehri}, {Gopinathan}, {Kim}, \&
  {Henning}}]{dib07}
{Dib}, S., {Shadmehri}, M., {Gopinathan}, M., {Kim}, J., \& {Henning}, T. 2007,
  ArXiv e-prints, 710

\bibitem[{{Elmegreen}(2004)}]{e04}
{Elmegreen}, B.~G. 2004, \mnras, 354, 367

\bibitem[{{Fukushige} \& {Heggie}(1995)}]{fh95}
{Fukushige}, T. \& {Heggie}, D.~C. 1995, \mnras, 276, 206

\bibitem[{{Gaburov} \& {Gieles}(2008)}]{gg08}
{Gaburov}, E. \& {Gieles}, M. 2008, ArXiv e-prints, 801

\bibitem[{{Heggie} \& {Mathieu}(1986)}]{hm85}
{Heggie}, D.~C. \& {Mathieu}, R.~D. 1986, in Lecture Notes in Physics, Berlin
  Springer Verlag, Vol. 267, The Use of Supercomputers in Stellar Dynamics, ed.
  P.~{Hut} \& S.~L.~W. {McMillan}, 233--+

\bibitem[{{Kroupa}(1995)}]{k95}
{Kroupa}, P. 1995, \mnras, 277, 1491

\bibitem[{{Kroupa}(2001)}]{k01}
{Kroupa}, P. 2001, \mnras, 322, 231

\bibitem[{{Kroupa}(2002)}]{k02}
{Kroupa}, P. 2002, Science, 295, 82

\bibitem[{{Kroupa} {et~al.}(2001){Kroupa}, {Aarseth}, \& {Hurley}}]{kah01}
{Kroupa}, P., {Aarseth}, S., \& {Hurley}, J. 2001, \mnras, 321, 699

\bibitem[{{Kroupa} {et~al.}(1991){Kroupa}, {Gilmore}, \& {Tout}}]{kgt91}
{Kroupa}, P., {Gilmore}, G., \& {Tout}, C.~A. 1991, \mnras, 251, 293

\bibitem[{{Lada} {et~al.}(1984){Lada}, {Margulis}, \& {Dearborn}}]{lmd84}
{Lada}, C.~J., {Margulis}, M., \& {Dearborn}, D. 1984, \apj, 285, 141

\bibitem[{{Larson}(1998)}]{l98}
{Larson}, R.~B. 1998, \mnras, 301, 569

\bibitem[{{Ma{\'{\i}}z Apell{\'a}niz} \& {{\'U}beda}(2005)}]{ma05}
{Ma{\'{\i}}z Apell{\'a}niz}, J. \& {{\'U}beda}, L. 2005, \apj, 629, 873

\bibitem[{{Makino} {et~al.}(2003){Makino}, {Fukushige}, {Koga}, \&
  {Namura}}]{mfkn03}
{Makino}, J., {Fukushige}, T., {Koga}, M., \& {Namura}, K. 2003, \pasj, 55,
  1163

\bibitem[{{Motte} {et~al.}(1998){Motte}, {Andre}, \& {Neri}}]{man98}
{Motte}, F., {Andre}, P., \& {Neri}, R. 1998, \aap, 336, 150

\bibitem[{{Motte} {et~al.}(2001){Motte}, {Andr{\'e}}, {Ward-Thompson}, \&
  {Bontemps}}]{maw01}
{Motte}, F., {Andr{\'e}}, P., {Ward-Thompson}, D., \& {Bontemps}, S. 2001,
  \aap, 372, L41

\bibitem[{{Murray} \& {Lin}(1996)}]{ml96}
{Murray}, S.~D. \& {Lin}, D.~N.~C. 1996, \apj, 467, 728

\bibitem[{{Pflamm-Altenburg} \& {Kroupa}(2006)}]{pk06}
{Pflamm-Altenburg}, J. \& {Kroupa}, P. 2006, \mnras, 373, 295

\bibitem[{{Sabbi} {et~al.}(2007){Sabbi}, {Sirianni}, {Nota}, {Tosi},
  {Gallagher}, {Smith}, {Angeretti}, {Meixner}, {Oey}, {Walterbos}, \&
  {Pasquali}}]{sabbi07}
{Sabbi}, E., {Sirianni}, M., {Nota}, A., {et~al.} 2007, ArXiv e-prints, 710

\bibitem[{{Subr} {et~al.}(2008){Subr}, {Kroupa}, \& {Baumgardt}}]{skb08}
{Subr}, L., {Kroupa}, P., \& {Baumgardt}, H. 2008, ArXiv e-prints, 801

\bibitem[{{Tilley} \& {Pudritz}(2005)}]{tp05}
{Tilley}, D.~A. \& {Pudritz}, R.~E. 2005, in Protostars and Planets V, 8473--+

\bibitem[{{Weidner} \& {Kroupa}(2004)}]{wk04}
{Weidner}, C. \& {Kroupa}, P. 2004, \mnras, 348, 187

\end{thebibliography}
\makeatletter   \renewcommand{\@biblabel}[1]{[#1]}   \makeatother

\label{lastpage}

\end{document}